\documentclass{article}
\usepackage{amsmath,amsfonts,setspace,cancel,url,hyperref,verbatim,color,caption}
\usepackage[a4paper]{geometry}
\usepackage[nosort]{cite}

\newcommand{\be}{\begin{equation}}
\newcommand{\ee}{\end{equation}}
\newcommand{\bba}{\bar{a}}
\newcommand{\bbb}{\bar{b}}
\newcommand{\bbc}{\bar{c}}
\newcommand{\bbd}{\bar{d}}
\newcommand{\bee}{\bar{e}}
\newcommand{\bbf}{\bar{f}}

\newcommand{\brho}{\bar{\rho}} 
\newcommand{\bfo}{\bar{4}}
\newcommand{\bfi}{\bar{5}}

\newcommand{\balpha}{\bar{\alpha}}
\newcommand{\bbeta}{\bar{\beta}}
\newcommand{\bgamma}{\bar{\gamma}}
\newcommand{\bdelta}{\bar{\delta}}
\newcommand{\TM}{M}
\newcommand{\TMP}{\tilde{M}}
\newcommand{\Six}{\xi}
\newcommand{\diag}{\textrm{diag}}
\numberwithin{equation}{section}

\newcommand\Tstrut{\rule{0pt}{3ex}}         
\newcommand\Bstrut{\rule[-1.3ex]{0pt}{0pt}}   

\newcommand{\ben}{\begin{displaymath}}
\newcommand{\een}{\end{displaymath}}
\newcommand{\bea}{\begin{eqnarray}}
\newcommand{\eea}{\end{eqnarray}}

\newcommand{\bean}{\begin{eqnarray*}}
\newcommand{\eean}{\end{eqnarray*}}
\newcommand{\beqs}{\begin{eqnarray}}
\newcommand{\eeqs}{\end{eqnarray}}

\begin{document}

\begin{titlepage}

\begin{center}

\hfill  LMU-ASC 64/15

\vskip 1.5cm

{\LARGE \sc Dualising consistent IIA / IIB truncations}

\vskip 1cm

{\large \sc Emanuel Malek$^*$ and Henning Samtleben$^\dagger$} \\

\vskip 25pt

{\em  $^*$ Arnold Sommerfeld Center for Theoretical Physics, \\
Department f\"ur Physik, Ludwig-Maximilians-Universit\"at M\"unchen,\\ Theresienstra{\ss}e 37, 80333 M\"unchen, Germany.\\[2ex]
$^\dagger$ Universit\'{e} de Lyon, Laboratoire de Physique, UMR 5672, CNRS, \\
\'{E}cole Normale Sup\'{e}rieure de Lyon, \\
46, all\'{e}e d'Italie, F-69364 Lyon cedex 07, France.}\\[2ex]

{{\tt E.Malek@lmu.de, \quad Henning.Samtleben@ens-lyon.fr\quad }} \\

\end{center}

\vskip 0.5cm

\begin{center} {\bf Abstract}\\[3ex]
\end{center}

\noindent 
We use exceptional field theory to establish a duality between certain consistent 7-dimensional truncations with maximal SUSY from IIA to IIB. We use this technique to obtain new consistent truncations of IIB on $S^3$ and $H^{p,q}$ and work out the explicit reduction formulas in the internal sector. We also present uplifts for other gaugings of 7-d maximal SUGRA, including theories with a trombone gauging. Some of the latter can only be obtained by a non-geometric compactification.

\thispagestyle{empty}

\end{titlepage}

\newpage
\tableofcontents 

\section{Introduction} \label{s:Intro}

The consistent Kaluza-Klein truncation of higher-dimensional (super)gravity to lower-dimensional theories is an old and generically difficult problem due to the highly non-linear gravitational field equations~\cite{Duff:1984hn}. Typically, consistent truncations require very particular backgrounds together with very particular matter couplings of the higher-dimensional theory, see e.g.~\cite{deWit:1986iy,Nastase:1999kf,Cvetic:2000dm}. Recent progress has come from the realisation of non-toroidal geometric compactifications via generalised Scherk-Schwarz-type compactifications on an extended spacetime within duality covariant formulations of the higher-dimensional supergravity theories~\cite{Aldazabal:2011nj,Geissbuhler:2011mx,Berman:2012uy,Aldazabal:2013mya,Lee:2014mla,Hassler:2014sba,Hohm:2014qga,Cho:2015lha}. In this language, finding consistent Kaluza-Klein reduction Ans\"atze translates into the search for Scherk-Schwarz twist matrices satisfying a number of differential consistency equations in the physical coordinates. Most recently, this has been used to work out the full Kaluza-Klein reduction for the AdS$_5\times S^5$ reduction of IIB supergravity in the framework of exceptional field theory~\cite{Baguet:2015sma}.

In this paper we use this framework to study consistent truncations from IIA and IIB supergravity down to seven dimensional gauged supergravities. Specifically, we establish a duality relating consistent IIA and IIB truncations for certain gaugings of maximal 7-dimensional supergravity. We then employ this duality to derive new consistent truncations of type IIB theory on the three sphere $S^3$, as well as on hyperboloids $H^{p,q}$, which lead to compact $SO(4)$, non-compact $SO(p,q)$ and non-semisimple $CSO(p,q,r)$ gaugings, respectively. Finally, we discuss new uplifts to type IIA / IIB of gauged supergravities involving gauging of the trombone scaling symmetry. In this final set of gaugings, we find that some can only be obtained by non-geometric compactifications\footnote{Here we refer to global non-geometry, where the structure group of the manifold is not contained within the geometric subgroup of the U-duality group.}, in a set-up reminiscent of that recently discussed in \cite{Shahbazi:2015sba}.

Let us get more specific about the 7-dimensional theories discussed in this paper. In general, the fluxes in half-maximal supergravity are parametrized by an antisymmetric tensor $X_{ABC}$ of the T-duality group $SO(d,d)$ \cite{Kaloper:1999yr}, which encodes the T-duality chain of \cite{Shelton:2005cf}
\bea
X_{ABC} &:&
H_{abc} \longrightarrow f_{ab}{}^c \longrightarrow Q_{a}{}^{bc} \longrightarrow R^{abc}
\;,
\eea
as well as two $SO(d,d)$ vectors $X_A$ and $f_A$, \cite{Bergshoeff:2007vb}, the latter of which encodes the trombone gaugings. Because the trombone symmetry is an on-shell symmetry, theories with non-zero $f_A$ can only be defined at the level of the equations of motion \cite{LeDiffon:2008sh}. For $d=3$, i.e.\ reduction to seven dimensions, $X_{ABC}$ splits into two irreducible representations
\bea
{\bf 20} &\longrightarrow& {\bf 10} + {\bf 10'}
\;,
\nonumber\\
X_{ABC} &=& 
\Gamma_{ABC}{}^{\alpha\beta} M_{\alpha\beta} +  \Gamma_{ABC\,\alpha\beta} \tilde{M}^{\alpha\beta}
\;,
\eea
with the $SO(3,3)$ $\Gamma$-matrices (or 't Hooft symbols, see for example appendix B of \cite{Dibitetto:2012rk}), and symmetric matrices $M_{\alpha\beta}$, $\tilde{M}^{\alpha\beta}$. Here the indices $\alpha, \beta = 1, \ldots 4$ are fundamental $SL(4) \simeq Spin(3,3)$ spinor indices. Similarly, the vectors can be written in terms of the $\mathbf{6}$ of $SL(4)$ as
\begin{equation}
 X_A = \frac{1}{2}\, \Gamma_A{}^{\alpha\beta} \xi_{\alpha\beta} \,, \qquad f_A = \frac12\Gamma_A{}^{\alpha\beta} \tau_{\alpha\beta} \,.
\label{the6s}
\end{equation}
For simplicity's sake we will take $X_A = f_A = 0$ for the following discussion although we will reintroduce them later on.

Depending on the choice of $M_{\alpha\beta}$, $\tilde{M}^{\alpha\beta}$, there are various one-parameter families of seven-dimensional gaugings most of which are of locally non-geometric origin~\cite{Dibitetto:2012rk}. A distinguished role is played by the theories satisfying the condition
\bea
M_{\alpha\beta} \tilde{M}^{\alpha\beta} &=& 0\;.
\eea
First, these can be consistently embedded into the maximal theory and second the subset where either $M_{\alpha\beta}$ or $\tilde{M}^{\alpha\beta}$ is non-degenerate allow for a geometric uplift to the type-I theory in ten dimensions as compactifications on the sphere $S^3$ and hyperboloids $H^{p,q}$. For the sphere case, the reduction formulas have been worked out in~\cite{Cvetic:2000dm} and later explained in the context of generalized geometry/double field theory~\cite{Dibitetto:2012rk,Lee:2014mla,Bosque:2015jda}.
The duality
\bea
M_{\alpha\beta} &\longleftrightarrow& \tilde{M}^{\alpha\beta}
\;,
\label{flip0}
\eea
is a symmetry of the quadratic constraints ensuring consistency of the gauging, as a manifestation of a particular triple T-duality~\cite{Dibitetto:2012rk,Blumenhagen:2014gva}, generated by an element of $O(3,3)$ rather than $SO(3,3)$.

In this paper, we will study the embedding of these structures in the maximal theory with U-duality group $SL(5)$. The above representations are embedded into U-duality representations according to
\bea
SO(3,3) &\subset& SL(5)\;,\nonumber\\
{\bf 10} &\subset& {\bf 15}\;,\nonumber\\
{\bf 10'} &\subset& {\bf 40'}
\;.
\label{TU}
\eea
Now the duality (\ref{flip0}) is no longer a symmetry of one and the same theory. Instead, the different embeddings (\ref{TU}) into the representations of the U-duality group induce inequivalent maximal seven-dimensional theories with gauge groups $CSO(p,q,1)$ for the IIA background and $SO(p,q)$ for the IIB background, respectively~\cite{Samtleben:2005bp}. These theories only coincide after truncation to the half-maximal sector. The IIA uplift has been given in \cite{Hohm:2014qga} via a generalized Scherk-Schwarz Ansatz in an exceptional space in the framework of exceptional field theory~\cite{Hohm:2013pua}. Here we realise the duality (\ref{flip0}) as an outer automorphism of $SL(4)$ acting on the Scherk-Schwarz twist matrices, and thereby derive the full IIB reduction Ansatz. In particular, the duality exchanges the IIA and their dual IIB coordinates within the 10 coordinates of the exceptional space~\cite{Berman:2010is,Blair:2013gqa}
\bea
{\bf 10} &\longrightarrow& {\bf 3}_{\rm IIA} +{\bf 3'}_{\rm IIB} + {\bf 3}+ {\bf 1}
\;.
\label{3331}
\eea
We will also show how the triple T-duality acting on the $\mathbf{6}$'s \cite{Dibitetto:2012rk}
\begin{equation}
 \xi_{\alpha\beta} \longleftrightarrow \xi^{\alpha\beta} = \frac{1}{2} \epsilon^{\alpha\beta\gamma\delta} \xi_{\gamma\delta} \,, \qquad \tau_{\alpha\beta} \longleftrightarrow \tau^{\alpha\beta} = \frac{1}{2} \epsilon^{\alpha\beta\gamma\delta} \tau_{\gamma\delta} \,,
\end{equation}
is realised in the maximal theory.

The paper is organized as follows. In section~2 we briefly review the pertinent structures of the relevant
exceptional field theory and its generalized Scherk-Schwarz reduction ansatz.
In section~3 we realize the duality (\ref{flip0}) on the Scherk-Schwarz twist matrix, relating consistent IIA / IIB  truncations.
As an application we work out the full truncation Ans\"atze for the internal sectors of the IIA and IIB reductions.
In particular, this establishes the consistency of the $S^3$ reduction of the IIB theory.
Finally, in section 5 we extend the analysis to the construction of more general twist matrices 
and obtain new uplifts of various maximal supergravities including those in which the trombone scaling symmetry is gauged.

\section{EFT and 7-dimensional maximal gauged SUGRA} \label{s:Review}

Our key tool for the study of consistent truncations is the `exceptional field theory' (EFT) \cite{Hohm:2013pua,Hohm:2013vpa,Hohm:2013uia,Godazgar:2014nqa} with its associated extended geometry, see~\cite{Berman:2010is,Coimbra:2011ky,Coimbra:2012af}. This is the duality covariant formulation of higher-dimensional supergravity which renders manifest the exceptional symmetry groups that are known to appear under dimensional reduction~\cite{Cremmer:1979up}.
The formulation of interest for studying reductions to maximal seven-dimensional supergravity, is the $SL(5)$ exceptional field theory. Apart from metric and scalars, it carries 10 vectors ${\cal A}_\mu{}^{ab}$, as well as 5 two-forms ${\cal B}_{\mu\nu\,a}$ and 5 three-forms ${\cal C}_{\mu\nu\rho}{}^a$, all fields depending on 7 external and 10 internal coordinates $\{ x^\mu, Y^{ab} \}$, $\mu=0, \dots, 6$; $a=1, \dots, 5$ with all fields subject to the section constraint~\cite{Berman:2011cg}
\bea
\partial_{[ab} \otimes \partial_{cd]} &\equiv& 0\;.
\label{section}
\eea
Three-forms enter the Lagrangian only under internal derivatives as $\partial_{ab} {\cal C}_{\mu\nu\rho}{}^b$. While the full $SL(5)$ exceptional field theory has not yet been worked out (see \cite{Abzalov:2015ega,Hohm:2015xna,Wang:2015hca} for EFTs in higher dimensions), its scalar sector has been given and studied in \cite{Berman:2010is,Park:2013gaj,Blair:2014zba}. The 14 scalar fields parametrize a unit-determinant symmetric $5\times5$ matrix ${\cal M}^{ab}$, i.e.\ form the coordinates of the coset space $SL(5)/SO(5)$\,.
W.r.t.\ the generalized space, ${SL}(5)$ generalized diffeomorphisms act according to
\bea
\delta V^a &=& \Lambda^{bc}\partial_{bc} V^a+2 \partial_{bc}\Lambda^{ab} V^c +
\frac25\,\partial_{bc}\Lambda^{bc}\,V^a
\;,
\nonumber\\
\delta V_a &=& \Lambda^{bc}\partial_{bc} V_a-2 \partial_{ab}\Lambda^{bc} V_c -
\frac25\,\partial_{bc}\Lambda^{bc}\,V_a
\;,
\eea
on weight zero tensors in the fundamental representations of $SL(5)$.
The section constraint (\ref{section}) admits two solutions \cite{Blair:2013gqa}. Breaking the U-duality group $SL(5)$ down to  the geometric $SL(3)$, the internal coordinates decompose into
\bea
Y^{ab} &\longrightarrow& \{
Y^{\alpha\beta}, Y^{\alpha5}
\}
~\longrightarrow~
\{ Y^{m4}, Y^{mn}, Y^{m5}, Y^{45} \}
\;,\qquad
\alpha=1, \dots, 4\,;\; m=1,2,3\;,
\label{Yex}
\eea
c.f.~(\ref{3331}),
and it is easy to see that the section constraint (\ref{section}) is 
satisfied by restricting the coordinate dependence of all fields onto
\bea
\left\{
\begin{array}{lc}
y^m\equiv Y^{m4} & ({\rm IIA})\\
\tilde{y}_m\equiv\frac12\,\epsilon_{mnp}Y^{np} & ({\rm IIB})
\end{array}
\right.
\;,
 \label{eq:Coords}
\eea
respectively. Depending on the higher-dimensional origin, it is convenient to parametrize the scalar matrix ${\cal M}^{ab}$ in a IIA or IIB basis according to
\bea
{\cal M}_{\rm IIA}^{ab} \!\!\!\!&=&\!\!\!\!\!\!
{\footnotesize
\left(\!\!\!\!
\begin{array}{ccc}
\!\!\!e^{\varphi/2}\,g^{2/5} g^{mn} \!+\! e^{-\varphi/2} g^{-3/5}\,B^mB^n \!\!\!
&\!\!\! e^{-\varphi/2} g^{-3/5}\,B^m
\!\!\!&\!\!\! - g^{2/5} g^{mk} e^{\varphi/2}\, C_k \!+\! e^{-\varphi/2}  g^{-3/5}\, C B^m
\\
e^{-\varphi/2}  g^{-3/5}\, B^m \!\!&\!\! e^{-\varphi/2} g^{-3/5}\, \!\!\!&\!\!\! e^{-\varphi/2} g^{-3/5}\,C \\
- g^{2/5} g^{mk} e^{\varphi/2} C_k \!+\! e^{-\varphi/2} g^{-3/5}\, C B^m \!\!\!&\!\!\!
e^{-\varphi/2}\,g^{-3/5}\,C \!\!&\!\!
e^{-\varphi/2} g^{-3/5}\, C^2 \!+\! g^{2/5} ( e^{-\varphi} \!+\!  e^{\varphi/2} g^{kl}C_kC_l)\!\!\!
\end{array}\!
\right)}\;,\nonumber\\
{\cal M}_{\rm IIB}^{ab} \!\!\!\!\!&=& \!\!\!\!\!
\left(
\begin{array}{cc}
g^{-3/5} g^{mn} &
-g^{-3/5}   g^{mn} \,C^v{}_{n}
\\
-g^{-3/5}   g^{nk} \,C^u{}_k &
g^{-3/5}   C^{u}{}_m g^{mn} C^v{}_n + g^{2/5}\,H^{uv}
\end{array}
\right)
\;, \label{eq:MParam}
\eea
where for IIA $g_{mn}$ is the metric, $C_m$ is the Ramond-Ramond one-form, $B^m = \frac{1}{2} \epsilon^{mnp} B_{np}$ is the dualised Kalb-Ramond two-form, $C = \frac{1}{3!} \epsilon^{mnp} C_{mnp}$ is the dualised Ramond-Ramond three-form and $\varphi$ is the dilaton. For IIB, we follow the conventions of \cite{Blair:2013gqa} so that all four-dimensional indices are placed ``upside-down''. Thus, $g^{mn}$ represents the metric, $C_{m}{}^{u} = \left(B_m, C_m\right) =\frac{1}{2} \epsilon_{mnp} \left( B^{np}, C^{np} \right)$ represents the $SL(2)$ doublet formed from the Kalb-Ramond and Ramond-Ramond two-forms and $H^{uv}$ is the $SL(2)$ matrix parameterised by the dilaton $\varphi$ and Ramond-Ramond scalar $C_0$ as follows
\begin{equation}
 H^{uv} = \begin{pmatrix}
  e^{\varphi} & - C_0 e^{\varphi} \\ - C_0 e^{\varphi} &  e^{\varphi} \left(C_0\right)^2 + e^{-\varphi}
 \end{pmatrix} \,.
\end{equation}
Throughout the paper the metric will be given in Einstein frame, unless otherwise specified.

The EFT formulation of supergravity is a powerful tool for the study of consistent truncations, since a number of geometrically non-trivial reductions can be reformulated as generalized Scherk-Schwarz reductions on the extended space~\cite{Aldazabal:2011nj,Geissbuhler:2011mx,Dibitetto:2012rk,Berman:2012uy,Musaev:2013rq,Aldazabal:2013mya,Lee:2014mla,Hassler:2014sba,Hohm:2014qga}. In the reduction Ansatz, all dependence on the internal coordinates is carried by an $SL(5)$ valued twist matrix $ U_{a}{}^{\bar{a}}(Y)$ with the scalar fields reducing according to
\bea
{\cal M}_{ab}(x,Y) &=& U_{a}{}^{\bar{a}}(Y)\,U_{b}{}^{\bar{b}}(Y)\,{M}_{\bar{a}\bar{b}}(x)\;, 
\label{ssscalars}
\eea
and the remaining EFT fields factorizing as \cite{Hohm:2014qga}
\bea
 G_{\mu\nu}(x,Y) &=& \rho^{-2}(Y)\,G_{\mu\nu}(x)\;,\nonumber\\
  {\cal A}_{\mu}{}^{ab}(x,Y) &=& \rho^{-1}(Y)\, A_{\mu}{}^{\bar{a}\bar{b}}(x)\,U_{\bar{a}\bar{b}}{}^{ab}(Y) \;, 
 \nonumber\\
  {\cal B}_{\mu\nu\,a}(x,Y) &=& \rho^{-2}(Y)\, B_{\mu\nu\,\bar{a}}(x)\,U_{a}{}^{\bar{a}}(Y)  \;,
 \nonumber\\
   {\cal C}_{\mu\nu\rho}{}^a(x,Y) &=& \rho^{-3}(Y)\, C_{\mu\nu\rho}{}^{\bar{a}}(x)\,U_{\bar{a}}{}^{a}(Y)  \;,
\label{ssforms}
\eea
with a scalar function $\rho(Y)$\,. The 7-dimensional metric of the full 10-dimensional type II theory, $g_{\mu\nu}$, is related to $G_{\mu\nu}$ above by
\begin{equation}
 g_{\mu\nu}(x,Y) = |g|^{-1/5} G_{\mu\nu}(x,Y) \,,
\end{equation}
where $|g|$ here is the determinant of the metric in the internal directions and $G_{\mu\nu}(x)$ is the metric of the 7-dimensional gauged SUGRA. Consistency of the reduction Ansatz translates into the set of differential equations \cite{Berman:2012uy} (we use the conventions of \cite{Hohm:2014qga})
\bea
\partial_{ab} U_{(\bar{a}}{}^a U_{\bar{b})}{}^{b}  &\stackrel{!}{=}&
-\, \rho \,S_{\bar{a}\bar{b}}
\;,\nonumber\\
\epsilon^{abcef} \left( U_{ef}{}^{\bar{a}\bar{b}}  \, \partial_{ab} U_c{}^{\bar{c}} 
-
U_{ef}{}^{[\bar{a}\bar{b}}\,\partial_{ab} U_c{}^{\bar{c}]} \right)
&\stackrel{!}{=}&
 2 \rho \, Z^{\bar{a}\bar{b},\bar{c}}  
\;,
\nonumber\\
  \partial_{cd} U_{\bar{a}\bar{b}}{}^{cd}  
- 6\,\rho^{-1}  \, U_{\bar{a}\bar{b}}{}^{cd} \,\partial_{cd} \rho &\stackrel{!}{=}& 
- 2 \rho\,\tau_{\bar{a}\bar{b}}
\;,
\label{consistent}
\eea
for the twist matrices, with $U_{ab}{}^{\bar{a}\bar{b}}\equiv U_{[a}{}^{\bar{a}}U_{b]}{}^{\bar{b}}$, and constant tensors $S_{\bar{a}\bar{b}}$, $Z^{\bar{a}\bar{b},\bar{c}}$, $\tau_{\bar{a}\bar{b}}$ transforming in the ${\bf 15}$, ${\bf 40'}$, and ${\bf 10}$, of $SL(5)$, respectively. These tensors form the torsion of the Weitzenb\"ock connection of EFT \cite{Coimbra:2011ky,Berman:2013uda,Blair:2014zba,Lee:2014mla} and correspond to the embedding tensors of maximal $D=7$ supergravity, which describe the allowed gaugings of the seven-dimensional theory~\cite{Samtleben:2005bp}.
The quadratic constraints which these tensors need to satisfy for consistency are a direct consequence of their definition by (\ref{consistent}) together with the section constraint (\ref{section}) and ensure that the gauge group closes. For later convenience, we spell out these equations
\begin{equation}
 \begin{split}
  S_{\bba\bbd} Z^{\bbd(\bbb,\bbc)} - \frac14\, \epsilon_{\bba\bbd\bee\bbf\bar{g}} Z^{\bbd\bee,\bbb}Z^{\bbf\bar{g},\bbc} + \frac13\,\tau_{\bba\bbd} Z^{\bbd(\bbb,\bbc)} &=-\frac19\, \delta_{\bba}{}^{(\bbb}\, \epsilon^{\bbc)\bbd\bee\bbf\bar{g}} \tau_{\bbd\bee}\tau_{\bbf\bar{g}} \,, \\
  S_{\bba\bbd}Z^{\bbb\bbc,\bbd} +\frac16\,\epsilon^{\bbb\bbc\bbd\bee\bbf}\,\tau_{\bee\bbf} S_{\bba\bbd} &= -\frac14 \, \delta_{\bba}{}^{[\bbb}\, \epsilon^{\bbc]\bbd\bee\bbf\bar{g}} \tau_{\bbd\bee}\tau_{\bbf\bar{g}} \,, \\
  S_{\bba\bbd}Z^{\bbb\bbc,\bbd} +\frac13\,\tau_{\bba\bbd}Z^{\bbb\bbc,\bbd} &= -\frac29\, \delta_{\bba}{}^{[\bbb}\, \epsilon^{\bbc]\bbd\bee\bbf\bar{g}} \tau_{\bbd\bee}\tau_{\bbf\bar{g}} \,.
 \end{split} \label{eq:QC}
\end{equation}
In particular, these identities imply that
\begin{equation}
 W^{\bba\bbb,\bbc} X_{\bba\bbb} = 0 \,,
\end{equation}
where
\begin{equation}
 W^{\bba\bbb,\bbc} = - Z^{\bba\bbb,\bbc} + \frac{1}{3} \epsilon^{\bba\bbb\bbc\bbd\bee} \tau_{\bbd\bee}\,, \label{eq:Intertwining}
\end{equation}
is the ``intertwining tensor'' coupling two-forms to the vector field strengths~\cite{deWit:2008ta} whose rank encodes the number of massive two-forms in the theory. The $X_{\bba\bbb}$ are the gauge generators evaluated in the vector representation,
which take the form
\begin{equation}
 \left(X_{\bba\bbb}\right)_{\bbc}{}^{\bbd} = \tau_{\bba\bbb,\bbc}{}^{\bbd} = \frac{1}{2} \epsilon_{\bba\bbb\bbc\bee\bbf} Z^{\bee\bbf,\bbd} + 2 \delta^{\bbd}_{[\bba} S_{\bbb]\bbc} + \frac{1}{3} \delta^{\bbd}_{\bbc} \tau_{\bba\bbb} + \frac{2}{3} \delta^{\bbd}_{[\bba} \tau_{\bbb]\bbc} \,,\label{eq:EmbeddingTensor}
\end{equation}
in terms of the embedding tensors (\ref{consistent}).
With $\tau_{\bar{a}\bar{b}}=0$, the corresponding theories are the conventional gaugings of $D=7$ supergravity constructed in~\cite{Samtleben:2005bp}. In particular, the gaugings triggered by $S_{\bar{a}\bar{b}}$ correspond to $CSO(p,q,5-p-q)$ gauge groups. The corresponding twist matrices for their $D=11$ embedding have been provided in \cite{Hohm:2014qga}. The gaugings triggered by $Z^{\bar{a}\bar{b},\bar{c}}$ contain theories with gauge groups $CSO(p,q,4-p-q) \times \left(U(1)\right)^{4-p-q}$ and IIB origin. We will construct the corresponding twist matrices in this paper.
A non-vanishing $\tau_{\bar{a}\bar{b}}$ corresponds to a gauging of the trombone scaling symmetry of the $D=7$ theory, resulting in a theory that can be defined on the level of the equations of motion but does not admit an action~\cite{LeDiffon:2008sh} while still allowing for an uplift to the IIA/IIB equations of motion.

\section{Dualising IIA / IIB truncations} \label{s:AutMot}

In the above we have reviewed how consistent truncations of the IIA/IIB theory are encoded in Scherk-Schwarz twist matrices on the extended space (\ref{3331}) satisfying the consistency conditions (\ref{consistent}) and the section constraint~(\ref{section}). In this section, we will first realize the duality (\ref{flip0}) on the twist matrices and the coordinates of extended space in order to map consistent IIA truncations into consistent IIB truncations. In particular, this will provide the full non-linear reduction Ans\"atze for the reduction of the IIB theory on $S^3$ and the hyperboloids $H^{p,q}$.

At the level of the effective seven-dimensional theories this duality is realized on the embedding tensors that define the maximal gaugings. Decomposing the embedding tensors under the T-duality group as $SL(5) \longrightarrow SL(4) \sim Spin(3,3)$ we find
\begin{equation}
 \begin{split}
  \mathbf{15} &\longrightarrow \mathbf{10} \oplus \mathbf{4} \oplus \mathbf{1} \,, \\
  {\mathbf{40'}} &\longrightarrow {\mathbf{20'}} \oplus {\mathbf{10'}} \oplus \mathbf{6} \oplus {\mathbf{4'}} \,, \\
  \mathbf{10} &\longrightarrow \mathbf{6} \oplus \mathbf{4} \,.
 \end{split}
 \label{split3}
\end{equation}
We will now discuss the $O(3,3)$ transformation (\ref{flip0}) that exchanges the $\mathbf{10} \longleftrightarrow {\mathbf{10}'}$ and maps the two $\mathbf{6}$'s into themselves. We will show that it corresponds to a duality between IIA and IIB truncations. This transformation extends $SL(4)\sim Spin(3,3)$ to $Pin(3,3)$, acting on $SL(4)$ as the outer automorphism.

\subsection{Duality as an outer automorphism} \label{s:Aut}
In order to consider type II truncations, we first perform a dimensional reduction of the exceptional space (\ref{3331}). In terms of $SL(4)$ irreducible representations, the coordinates decompose as $Y^{ab} \longrightarrow \left( Y^{\alpha\beta}, Y^{\alpha 5} \right)$, where $\alpha = 1, \ldots 4$, c.f.~(\ref{Yex}). We assume no dependence on the $Y^{\alpha 5}$, i.e.\ reduce the exceptional
space to the doubled space of DFT~\cite{Siegel:1993th,Hull:2009zb,Hohm:2010jy,Hohm:2010pp}, 
see \cite{Thompson:2011uw}.
Depending on the choice of the physical coordinates among the remaining $Y^{\alpha\beta}$, the theory is of
IIA or IIB origin according to (\ref{eq:Coords}).
Let us also introduce the notation
\begin{equation}
Y_{\alpha\beta} = \frac12 \epsilon_{\alpha\beta\gamma\delta} Y^{\gamma\delta}\;,\qquad
 \partial^{\alpha\beta} = \frac{1}{2} \epsilon^{\alpha\beta\gamma\delta} \partial_{\gamma\delta} \,,
\end{equation}
where $\epsilon^{\alpha\beta\gamma\delta}$ is the 4-dimensional totally-antisymmetric symbol. 
The flip between IIA and IIB coordinates in (\ref{3331}) is then realized as
\bea
Y_{\alpha\beta}  \longleftrightarrow Y^{\alpha\beta}\;.
\label{flipCoordinates}
\eea

We start from the following 
$GL(4)$ Ansatz for the $SL(5)$ Scherk-Schwarz twist matrix
\begin{equation}
 U_a{}^{\bba} = \begin{pmatrix}
  \omega^{-1/2} \, V_\alpha{}^{\balpha} & 0 \\ 0 & \omega^2
 \end{pmatrix} \,, \label{eq:DiagonalTwist}
\end{equation}
with $V_\alpha{}^{\balpha} \in SL(4)$. It follows from (\ref{consistent}) that this Ansatz can only produce gaugings in the $\mathbf{10}$'s and $\mathbf{6}$'s according to the decomposition of (\ref{split3}). 
The corresponding embedding tensors are given in terms of the twist by
\begin{equation}
 \begin{split}
  S_{\balpha\bbeta} &= \TM_{\balpha\bbeta} \equiv \rho^{-1} \omega V_{(\balpha}{}^\alpha \partial_{|\alpha\beta|} V_{\bbeta)}{}^\beta \,, \qquad Z^{\bfi(\balpha,\bbeta)} = \TMP^{\balpha\bbeta} \equiv \rho^{-1} \omega V_\alpha{}^{(\balpha} \partial^{|\alpha\beta|} V_\beta{}^{\bbeta)} \,, \\
  2 \tau_{\balpha\bbeta} &= - \rho^{-1} \omega \left( \partial_{\alpha\beta} V_{\balpha\bbeta}{}^{\alpha\beta} -5 V_{\balpha\bbeta}{}^{\alpha\beta} \partial_{\alpha\beta} \ln \omega + 6 V_{\balpha\bbeta}{}^{\alpha\beta} \partial_{\alpha\beta} \ln \left(\rho^{-1}\omega\right) \right)\,, \\
  6Z^{\bfi[\balpha,\bbeta]} &= 2\Six^{\balpha\bbeta} \equiv \rho^{-1} \omega \left( \partial^{\alpha\beta} V_{\alpha\beta}{}^{\balpha\bbeta} - 5 V_{\alpha\beta}{}^{\balpha\bbeta} \partial^{\alpha\beta} \ln \omega \right) \,, \label{eq:CC10+6}
 \end{split}
\end{equation}
where we use $V_{\alpha\beta}{}^{\balpha\bbeta} = V_{[\alpha}{}^{\balpha} V_{\beta]}{}^{\bbeta}$.
The explicit form of these equations shows that combining the flip (\ref{flipCoordinates}) with the 
$\mathbb{Z}_2$ outer automorphism of $SL(4)$ 
\begin{equation}
 V_\alpha{}^{\balpha} \longleftrightarrow \left(V^{-T}\right){\!}_{\balpha}{}^{\alpha} \,, 
 \label{flipV}
\end{equation}
induces the duality (\ref{flip0}) on the embedding tensor. 
 Concretely this takes 
\begin{equation}
 \TM_{\balpha\bbeta} \longleftrightarrow \TMP^{\balpha\bbeta} \,, \qquad \tau_{\balpha\bbeta} \longleftrightarrow \tau^{\balpha\bbeta} \,, \qquad \Six^{\balpha\bbeta} \longleftrightarrow \Six_{\balpha\bbeta} \,, \label{eq:Aut}
\end{equation}
where $\tau^{\balpha\bbeta} = \frac{1}{2} \epsilon^{\balpha\bbeta\bgamma\bdelta} \tau_{\bgamma\bdelta}$ and $\Six_{\balpha\bbeta} = \frac{1}{2} \epsilon_{\balpha\bbeta\bgamma\bdelta} \Six^{\bgamma\bdelta}$. Additionally, the dualisation of the coordinates (\ref{flipCoordinates}) exchanges IIA and IIB sections so that this duality relates IIA and IIB truncations. The NS-NS sector remains invariant under the duality since in the half-maximal theory both ${\bf 10}$ and ${\bf 10'}$ lie in the same $O(3,3)$ orbit~\cite{Dibitetto:2012rk}. Within the maximal theory the duality (\ref{flip0}) relates consistent truncations of the maximal theories, which in general have different gauge groups, vacua, and fluctuations.

The gaugings above are the only ones that survive the $\mathbb{Z}_2$ projection to the half-maximal theory \cite{Dibitetto:2012rk}.
In section \ref{s:4s}, we will also discuss gaugings which do not survive the $\mathbb{Z}_2$ projection (these are the $\mathbf{20}'$, $\mathbf{4}$'s and $\mathbf{1}$)  and we will show that the duality above does not, in general, hold for these cases.

\subsection{Example: IIA and IIB on $S^3$ and $H^{p,q}$} \label{s:S3}

Before discussing the duality further, let us apply it to work out the 
consistent truncation of IIB SUGRA on $S^3$ and on warped $H^{p,q}$ manifolds. 
According to the above discussion, these are dual to the consistent truncations of 
IIA on these manifolds and yield a gauging in the ${\mathbf{10}'} \subset{\mathbf{40}'}$ with gauge group $CSO(p,q,r)\times U(1)^{r}$ where $p + q + r = 4$.

Let us begin by reading off $\omega$ and $V_\alpha{}^{\balpha}$ from the IIA twist matrices for $CSO(p,q,r)$ gaugings given in \cite{Hohm:2014qga}. Here and throughout this paper we will order the rows and columns of $V_\alpha{}^{\balpha}$ as $\left(i, 4, x\right)$ with $i, j,  = 1, \ldots, 4-r$ and $x, y = 5-r, \ldots, 3$. The twist is
\begin{equation}
 V_\alpha{}^{\balpha} = \begin{pmatrix}
 \left(1-v\right)^{1/4} \delta_{i}{}^{\bar{\imath}} & - \left(1-v\right)^{-1/4} \eta_{ij} y^{j} & 0 \\
 - \left(1-v\right)^{-1/4} K \eta^{\bar{\imath}\bar{\jmath}} y_{\bar{\jmath}} & \left(1-v\right)^{-3/4} \left(1 + Ku\right) & 0 \\
 0 & 0 & \left(1-v\right)^{1/4} \mathbb{I}_{r}
 \end{pmatrix} \,, \quad 
 \omega = \left(1-v\right)^{1/10} \;, \label{eq:IIASphere}
\end{equation}
with $u = \delta_{ij} y^i y^j$, $v = \eta_{ij} y^i y^j$, and the matrix $\eta_{ij}$ being the $SO(p-1,q)$ invariant diagonal matrix. Moreover, $y^{i} = Y^{i 4}$ and we make no distinction between un/barred and upper/lower indices on the IIA coordinates $y^m$. From (\ref{eq:CC10+6}), we further see that setting $\rho=\omega$ induces vanishing trombone parameter $\tau_{\bar{a}\bar{b}}$ as required for these gaugings. Together, the twist matrix then induces the gauging
\begin{equation}
 \TM_{\bar{\imath}\bar{\jmath}} = \eta_{\bar{\imath}\bar{\jmath}}\,, \qquad \TM_{\bfo\bfo} = 1 \,, \qquad \TM_{\bar{x}\bar{y}} = 0 \,.
\end{equation}
The function $K(u,v)$ appearing in the twist  satisfies the differential equation
\begin{equation}
 2 \left(1-v\right) \left( u\, \partial_v K + v\, \partial_u K \right) = \left(\left(1+q-p\right) \left(1-v\right) - u\right) K - 1 \,,
 \label{diffK}
\end{equation}
with $u = \delta_{ij} y^i y^j$. This can be solved analytically for all allowed values $p, q$. The internal space corresponding to these truncations are warped hyperboloids $H^{p,q}$ together with $r$ flat directions \cite{Hohm:2014qga}.

We now apply the duality (\ref{flipCoordinates}), (\ref{flipV}), to obtain the IIB truncations on $H^{p,q}$ 
which give rise to the $CSO(p,q,r)$ gaugings in the ${\mathbf{10}'} \subset {\mathbf{40}'}$  
such that
\begin{equation}
 \TMP^{\bar{\imath}\bar{\jmath}} = \eta^{\bar{\imath}\bar{\jmath}} \,, \qquad \TMP^{\bfo\bfo} = 1 \,, \qquad \TMP^{\bar{x}\bar{y}} = 0 \,,
 \label{gM}
\end{equation}
c.f.~(\ref{eq:Aut}).
The IIB twist matrices are thus
\begin{equation}
 V_\alpha{}^{\balpha} = \begin{pmatrix}
 \left(1-\tilde{v}\right)^{-1/4} \left( \delta_{i}{}^{\bar{\imath}} + K \eta_{ij} \eta^{\bar{\imath}\bar{\jmath}} \tilde{y}^j \tilde{y}_{\bar{\jmath}} \right) & \left(1-\tilde{v}\right)^{1/4} K \eta_{ij} \tilde{y}^{j} & 0 \\
 \left(1-\tilde{v}\right)^{1/4} \eta^{\bar{\imath}\bar{\jmath}} \tilde{y}_{\bar{\jmath}} & \left(1-\tilde{v}\right)^{3/4} & 0 \\
 0 & 0 & \left(1-\tilde{v}\right)^{-1/4} \mathbb{I}_{r}
 \end{pmatrix} \,, \quad \omega = \left(1-\tilde{v}\right)^{1/10} , \label{eq:IIBSphere}
\end{equation}
with $\rho = \omega$ and where now $\tilde{y}_i$ are IIB coordinates \eqref{eq:Coords}, $\tilde{u} = \delta^{ij} \tilde{y}_i \tilde{y}_j$ and $\tilde{v}=\eta^{ij} \tilde{y}_i \tilde{y}_j$. Using \eqref{ssscalars} and the parameterisation \eqref{eq:MParam} we can read off the internal space of the compactification. At the origin of the scalar coset, $M_{\bba\bbb}(x) = \delta_{\bba\bbb}$, we find the background, given by:
\begin{equation}
 \begin{split}
  \mathring{ds}^2 &= \left(1+\tilde{u}-\tilde{v}\right)^{-3/4} \left[ \left( \delta^{ij} + \frac{ \eta^{ik} \tilde{y}_k \eta^{jl} \tilde{y}_l  }{1-\tilde{v}} \right) d\tilde{y}_i d\tilde{y}_j + \delta^{xy} d\tilde{y}_x d\tilde{y}_y \right] \\
  &\quad + \left(1+\tilde{u}-\tilde{v}\right)^{1/4} ds_7^2 \,, \\
  \mathring{B}^{mn} &= \left(1-\tilde{v}\right)^{-1/2} \left(\frac{1}{1+\tilde{u}-\tilde{v}} + K \right) \epsilon^{mnp} \eta_{pq} \tilde{y}^q \,, \\
  e^{\mathring{\varphi}} &= \left(1+\tilde{u}-\tilde{v}\right)^{-1/2} \,. \label{eq:IIBSphereInt}
 \end{split}
\end{equation}
Here we let $m, n = 1, 2, 3$ and we denote by $\mathring{B}^{mn}$ the Kalb-Ramond form and by $\mathring{\varphi}$ the dilaton. We recall that following the conventions of \cite{Blair:2013gqa} and matching the indices of the IIB coordinates \eqref{eq:Coords}, the four-dimensional IIB indices are ``upside-down'' compared to the usual placement. The internal space here is the a warped product $H^{p,q} \times \mathbb{R}^{r}$, where $H^{p,q}$ is the surface satisfying $\eta^{ij} \tilde{y}_i \tilde{y}_j + z^2 = 1$ in $\mathbb{R}^{4-r}$, with $z$ an additional coordinate. This coincides with the IIA background for this truncation, see \cite{Hohm:2014qga}. The Kalb-Ramond background field strength is given by
\begin{equation}
 \mathring{F}^{mnp} = 3 \partial^{[m} \mathring{B}^{np]} = \frac{\epsilon^{mnp}}{\left(1-\tilde{v}\right)^{1/2}\left(1+\tilde{u}-\tilde{v}\right)^2} \left( p - q - 2 + \left(\tilde{u}-\tilde{v}\right) \left(p-q\right) \right) \,,
\end{equation}
upon using (\ref{diffK}).

Using \eqref{ssscalars}, we can furthermore determine the full truncation Ansatz for the internal fields as fluctuations about the background \eqref{eq:IIBSphereInt}. To simplify the notation, we will for this discussion not distinguish between barred and un-barred indices and we will simply refer to the IIB coordinates as $y_i$, i.e.\ drop the tilde. Let us start by considering the case where $p + q = 4$. The truncation Ansatz can be elegantly formulated in terms of the harmonics
\begin{equation}
 {\cal Y}_\alpha = \left( y_m, \left(1-v\right)^{-1/2} \right) \,, \qquad {\cal Y}^\alpha = \eta^{\alpha\beta} {\cal Y}_\beta \,,
 \label{Y}
\end{equation}
the auxiliary metric
\begin{equation}
 \tilde{g}^{ij} = \eta^{ij} + \frac{\eta^{ik} y_k \eta^{jl} y_j}{1-\tilde{v}} \,, \qquad \tilde{g}_{ij} = \eta_{ij} - y_i y_j \,,
\end{equation}
with volume form
\begin{equation}
 \tilde{\omega}^{ijk} = \left(1-v\right)^{-1/2} \epsilon^{ijk} \,,
\end{equation}
and the auxiliary two form
\bea
\tilde{B}^{ij} = \tilde{\omega}^{ijk} \left( K \eta_{kl} y^l + y_k \right)\quad
\Longrightarrow\quad
3\, \partial^{[i} \tilde{B}^{jk]} = 2\, \tilde{\omega}^{ijk}\;,
\eea
see section 3 of \cite{Baguet:2015sma} on details of the construction. We note that only for the sphere case, when $\eta_{ij}=\delta_{ij}$, these auxiliary structures coincide with the background (\ref{eq:IIBSphereInt}). Furthermore, it will be useful to decompose the scalar fields ${M}_{ab}(x)$ as
\begin{equation}
 {M}_{ab} = \begin{pmatrix}
  \kappa^{-1} \left( m_{\alpha\beta} + m_{\alpha 5} m_{\beta 5} \right) & \kappa^{3/2} m_{\alpha 5} \\ \kappa^{3/2} m_{\beta 5} & \kappa^4
 \end{pmatrix} \,, \label{eq:ScalarAnsatz}
\end{equation}
with $SL(4)$ matrix $m_{\alpha\beta}$. The truncation formulae for the internal components of all IIB fields are then read off from \eqref{eq:MParam}, \eqref{ssscalars} and yield
\begin{equation}
 \begin{split}
  g^{ij} &= \kappa^{-5/4} \Delta^{3/5} \partial^i {\cal Y}_\alpha \partial^j {\cal Y}_\beta m^{\alpha\beta} \,, \\
  H_{uv} &= \Delta^{2/5} \begin{pmatrix}
   \kappa^{-5/2} {\cal Y}^\alpha {\cal Y}^\beta \left( m_{\alpha\beta} + m_{\alpha 5} m_{\beta 5} \right) & {\cal Y}^\alpha m_{\alpha 5} \\ {\cal Y}^\beta m_{\beta 5} & \kappa^{5/2}
  \end{pmatrix} \;,
  \\
  C^{ij,u} &= \left\{ \tilde{B}^{ij} ,\, 0 \right\} - \frac{2}{5} \tilde{\omega}^{ijk} \tilde{g}_{kl} \left\{ \Delta^{-1} \partial^l \Delta , \, - 5 \kappa^{-5/2} \Delta^{4/5} \partial^{l} {\cal Y}^{[\alpha} {\cal Y}^{\beta]} {\cal Y}^\gamma m_{\alpha 5} m_{\beta\gamma} \right\} \,, \\
  g_{\mu\nu} &= \kappa^{3/4} \Delta^{-1/5} G_{\mu\nu}(x) \,,
 \end{split}
\end{equation}
in terms of the objects (\ref{Y})--(\ref{eq:ScalarAnsatz}) and with the function $\Delta$ given by
\begin{equation}
 \Delta = \left( \mathcal{Y}^\alpha \mathcal{Y}^\beta m_{\alpha\beta} \right)^{-5/4} \,.
\end{equation}
It is straightforward to verify that at the scalar origin $M_{ab}(x)=\delta_{ab}$, these formulae reduce
to the background  (\ref{eq:IIBSphereInt}).

Let us now compare this result to the IIA truncation formulae on the dual background. Define, now in terms 
of the IIA coordinates, the harmonics
\begin{equation}
 {\cal Y}^\alpha = \left( y^i, \left(1-v\right)^{-1/2} \right) \,, \qquad {\cal Y}_\alpha = \eta_{\alpha\beta} {\cal Y}^\beta \,,
\end{equation}
the auxiliary metric (as before but now with the reverse position of indices)
\begin{equation}
 \tilde{g}_{ij} = \eta_{ij} + \frac{\eta_{ik} y^k \eta_{jl} y^l}{1-v} \,, \qquad \tilde{g}^{ij} = \eta^{ij} - y^i y^j \,,
\end{equation}
with volume form
\begin{equation}
 \tilde{\omega}_{ijk} = \left(1-v\right)^{-1/2} \epsilon_{ijk} \,,
\end{equation}
and the auxiliary two-form
\bea
\tilde{B}_{ij} &=& \tilde{\omega}_{ijk} \left( K \eta^{kl} y_l + y^k \right) 
\quad
\Longrightarrow\quad
3\, \partial_{[i} \tilde{B}_{jk]} = 2\, \tilde{\omega}_{ijk}
\;.
\eea
With the same scalar matrix \eqref{eq:ScalarAnsatz},
the truncation formulae for the internal components of all IIA fields are again read off from \eqref{eq:MParam}, \eqref{ssscalars} and yield
\begin{equation}
 \begin{split}
 g_{ij} &= \kappa^{-5/4} \Delta^{3/5} \partial_i {\cal Y}^\alpha \partial_j {\cal Y}^\beta m_{\alpha\beta} \,, \\
 e^{\varphi} &= \kappa^{5/2} \Delta^{2/5} \,, \\
 C_i &= \kappa^{-5/2} \partial_i \left({\cal Y}^\alpha m_{\alpha 5} \right) \,, \\
 B_{ij} &= \tilde{B}_{ij} - \frac{2}{5} \Delta^{-1} \tilde{\omega}_{ijk} \tilde{g}^{kl} \partial_l \Delta \,, \\
 C_{ijk} &= - \kappa^{-3/2} \Delta^{4/5} \tilde{\omega}_{ijk} m_{\alpha 5} m^{\alpha\beta} {\cal Y}_\beta \,, \\
 g_{\mu\nu} &= \kappa^{3/4} \Delta^{-1/5} G_{\mu\nu}(x) \,,
 \end{split}
\end{equation}
in terms of the above objects and with
\begin{equation}
 \Delta = \left(m^{\alpha\beta} {\cal Y}_\alpha {\cal Y}_\beta \right)^{-5/4}  \,.
\end{equation}
We can now see that the full reduction formulae of the IIA and IIB truncations coincide for the NS-NS sector and are related by the same $SL(4)$ outer automorphism we have used for the twists,
extended to the scalar fields (\ref{eq:ScalarAnsatz})
\begin{equation}
 m^{\alpha\beta} \longleftrightarrow m_{\alpha\beta} \,, \qquad {\cal Y}_\alpha \longleftrightarrow {\cal Y}^\alpha \,.
\end{equation}

Finally, let us also give the reduction formulae when $2 \leq p + q \leq 4$. To keep the notation more compact it will now be useful to use the dualised form potentials. For IIB these are
\begin{equation}
 C_{m}{}^u = \frac{1}{2} \epsilon_{mnp} C^{np,u} \,,
\end{equation}
while for IIA they are
\begin{equation}
 B^m = \frac{1}{2} \epsilon^{mnp} B_{np} \,, \qquad C = \frac{1}{3!} \epsilon^{mnp} C_{mnp} \,.
\end{equation}
Let us once again start with the IIB reduction. Recall that our convention is that $m = \left(i, x\right)$ where $\eta^{xy} = 0$ and $\eta^{ij} \neq 0$. Let
\begin{equation}
 \begin{split}
  {\cal Y}_\alpha &= \left( y_i ,\, \left(1-v\right)^{1/2}\!\!\! ,\, y_x \right) \,, \qquad {\cal Y}^\alpha = \eta^{\alpha\beta} {\cal Y}_\beta \,, \\
  \Delta &= \left( m_{\alpha\beta} {\cal Y}^\alpha {\cal Y}^\beta \right)^{-5/4} \,, \qquad \tilde{B}_i = \frac{1}{2} \epsilon_{ijk} \tilde{B}^{jk} = \left(1-v\right)^{-1/4} \left( K \eta_{ij} y^j + y_i \right) \,,
 \end{split}
\end{equation}
and $\tilde{g}_{ij}$, $\tilde{g}^{ij}$ as before. Then we obtain the IIB truncation formulae
\begin{equation}
 \begin{split}
  g^{mn} &= \kappa^{-5/4} \Delta^{3/5} \partial^m {\cal Y}_\alpha \partial^n {\cal Y}_\beta m^{\alpha\beta} \,, \\
  H_{uv} &= \Delta^{2/5} \begin{pmatrix}
  \kappa^{-5/2} {\cal Y}^\alpha {\cal Y}^\beta m_{\alpha\beta} & {\cal Y}^\alpha m_{\alpha 5} \\ {\cal Y}^\beta m_{\beta 5} & \kappa^{5/2}
  \end{pmatrix} \\
  C_{i}{}^u &= \left\{ \tilde{B}_i ,\, 0 \right\} - \frac{2}{5} \left(1-v\right)^{-1/2} \tilde{g}_{ij} \left\{ \Delta^{-1} \partial^j \Delta , \, - 5 \kappa^{-5/2} \Delta^{4/5} \partial^{j} {\cal Y}^{[\alpha} {\cal Y}^{\beta]} {\cal Y}^\gamma m_{\alpha 5} m_{\beta\gamma} \right\} \,, \\
  C_x{}^u &= \left(1-v\right)^{-1/2} \Delta^{4/5} \left\{ m_{x\alpha} {\cal Y}^\alpha, \, \kappa^{-5/2} \left( m_{\alpha\beta} m_{x5} - m_{\alpha 5} m_{x\beta} \right) {\cal Y}^\alpha {\cal Y}^\beta \right\} \,, \\
  g_{\mu\nu} &= \kappa^{3/4} \Delta^{-1/5} G_{\mu\nu}(x) \,.
 \end{split}
\end{equation}

The corresponding IIA formulae can be given in terms of
\begin{equation}
 \begin{split}
 {\cal Y}^\alpha &= \left( y^i,\, \left(1-v\right)^{1/2} ,\, y^x \right) \,, \qquad {\cal Y}_\alpha = \eta_{\alpha\beta} {\cal Y}^\beta \,, \\
 \Delta &= \left( m^{\alpha\beta} {\cal Y}_\alpha {\cal Y}_\beta \right)^{-5/4} \,, \qquad \tilde{B}^i = \left(1-v\right)^{-1/4} \left(K \eta^{ij} y_j + y^i \right) \,.
 \end{split}
\end{equation}
and read
\begin{equation}
 \begin{split}
 g_{mn} &= \kappa^{-5/4} \Delta^{3/5} \partial_m {\cal Y}^\alpha \partial_n {\cal Y}^\beta m_{\alpha\beta} \,, \\
 e^{\varphi} &= \kappa^{5/2} \Delta^{2/5} \,, \\
 C_m &= \kappa^{-5/2} \partial_m \left({\cal Y}^\alpha m_{\alpha 5} \right) \,, \\
 B^i &= \tilde{B}^{i} - \frac{2}{5} \left(1-v\right)^{-1/2} \Delta^{-1} \tilde{g}^{kl} \partial_l \Delta \,, \\
 B^x &= - \left(1-v\right)^{-1/2} \kappa \Delta^{4/5} m^{x\alpha} {\cal Y}_\alpha \,, \\
 C &= - \left(1-v\right)^{-1/2} \kappa^{-3/2} \Delta^{4/5} m_{\alpha 5} m^{\alpha \beta} {\cal Y}_\beta \,, \\
 g_{\mu\nu} &= \kappa^{3/4} \Delta^{-1/5} G_{\mu\nu}(x) \,.
 \end{split}
\end{equation}

\subsection{A no-go theorem
on IIA/IIB uplifts} \label{s:NoGo10s}

We have just shown that the IIA truncations with gaugings in the $\mathbf{10} \subset \mathbf{15}$ induce dual IIB truncations with gaugings in the ${\mathbf{10}'} \subset {\mathbf{40}'}$, according to the embedding (\ref{split3}). A natural question to ask is whether it is possible to obtain the $\mathbf{10} \subset \mathbf{15}$ gauging by a IIB truncation -- or equivalently the ${\mathbf{10}'} \subset {\mathbf{40}'}$ by a IIA truncation. We will now show that this cannot be done by analysing the symmetries of the embedding tensor.

In order to use symmetry properties of the embedding tensor, we work in the 10-dimensional representation with
\begin{equation}
 \tau_{\bba\bbb,\bbc\bbd}{}^{\bee\bbf} = 2\, \tau_{\bba\bbb,[\bbc}{}^{[\bee} \delta_{\bbd]}{}^{\bbf]} \,,
\end{equation}
where $\tau_{\bba\bbb,\bbc}{}^{\bbd}$ represents the embedding tensor as given in \eqref{eq:EmbeddingTensor}. The consistency equations \eqref{consistent} in this representation can conveniently be computed in terms of 
\bea
E_{\bba\bbb}{}^{ab} &\equiv& \rho^{-1}\,
U_{[\bba}{}^{a}
U_{\bbb]}{}^{b}
\;,
\eea
via the generalized Lie derivative
\begin{equation}
 \mathcal{L}_{E_{\bba\bbb}} E_{\bbc\bbd}{}^{ef}  = \frac{1}{2} E_{\bba\bbb}{}^{ab} \partial_{ab} E_{\bbc\bbd}{}^{ef} + \frac{1}{2} E_{\bbc\bbd}{}^{ef} \partial_{ab} E_{\bba\bbb}{}^{ab} + 2 E_{\bbc\bbd}{}^{a[e} \partial_{ab} E_{\bba\bbb}{}^{f]b} 
 \equiv - \tau_{\bba\bbb,\bbc\bbd}{}^{\bee\bbf} E_{\bee\bbf}{}^{ef}
 \,. \label{eq:10CC}
\end{equation}
We first assume that the twist only depends on IIA coordinates $Y^{m 4}$ and introduce the following notation
\begin{equation}
 E_{\bba\bbb}{}^{45} = R_{\bba\bbb} \,, \qquad E_{\bba\bbb}{}^{m 4} = K_{\bba\bbb}{}^m \,, \qquad E_{\bba\bbb}{}^{m 5} = \frac{1}{2} \epsilon^{mnp} L_{np,\bba\bbb} \,, \qquad E_{\bba\bbb}{}^{mn} = \epsilon^{mnp} T_{p,\bba\bbb} \,.
\end{equation}
Then the consistency condition (\ref{eq:10CC}) takes the form
\begin{equation}
 \begin{split}
  - \tau_{\bba\bbb,\bbc\bbd}{}^{\bee\bar{f}} K_{\bee\bar{f}}{}^{m} &= {\cal L}_{\bba\bbb} K_{\bbc\bbd}{}^{m} \equiv K_{\bba\bbb}{}^{n } \partial_n K_{\bbc\bbd}{}^{m} - K_{\bbc\bbd}{}^{n} \partial_n K_{\bba\bbb}{}^{m} \,, \\
  - \tau_{\bba\bbb,\bbc\bbd}{}^{\bee\bar{f}} R_{\bee\bar{f}} &= K_{\bba\bbb}{}^{m} \partial_m R_{\bbc\bbd} - K_{\bbc\bbd}{}^{m} \partial_m R_{\bba\bbb} \,, \\
  - \tau_{\bba\bbb,\bbc\bbd}{}^{\bee\bbf} L_{mn,\bee\bbf} &= {\cal L}_{\bba\bbb} L_{mn,\bbc\bbd} - 3 K_{\bba\bbb}{}^{p} \partial_{[p} L_{mn],\bba\bbb} - 2 T_{\bbc\bbd[m} \partial_{n]} R_{\bba\bbb} + 2 R_{\bbc\bbd} \partial_{[m} T_{|\bba\bbb|,n]} \,, \\
  - \tau_{\bba\bbb,\bbc\bbd}{}^{\bee\bbf} T_{m,\bee\bbf} &= {\cal L}_{\bba\bbb} T_{m,\bbc\bbd} - 2 K_{\bbc\bbd}{}^n \partial_{[n} T_{m],\bba\bbb} \,. \label{eq:10CCIIA}
 \end{split}
\end{equation}
Here ${\cal L}_{\bba\bbb}$ denotes the standard Lie derivative with diffeomorphism parameter $K_{\bba\bbb}{}^m$ acting on vectors $K_{\bba\bbb}{}^m$ and co-vectors $T_{m, \bba\bbb}$ and two-forms $L_{mn, \bba\bbb}$. Note that in the first two equations, the right-hand side is antisymmetric under the exchange of $[\bba \bbb] \leftrightarrow [\bbc\bbd]$. Thus, we see that certain contractions of the symmetric part of the embedding tensor vanish. Its symmetric part is given by the intertwining tensor \eqref{eq:Intertwining},
\begin{equation}
 \tau_{\bba\bbb,\bbc\bbd}{}^{\bee\bbf} + \tau_{\bbc\bbd,\bba\bbb}{}^{\bee\bbf} = \epsilon_{\bba\bbb\bbc\bbd\bar{g}} W^{\bee\bbf,\bar{g}} = - \epsilon_{\bba\bbb\bbc\bbd\bar{g}} Z^{\bee\bbf,\bar{g}} + 8\,\delta^{\bee\bbf\bar{g}\bar{h}}_{\bba\bbb\bbc\bbd} \tau_{\bar{g}\bar{h}} \,. \label{eq:SymTorsion}
\end{equation}
Thus, a necessary requirement for gaugings to be lifted to IIA is that
\begin{equation}
 3 Z^{\bbb\bbc,\bba} K_{\bbb\bbc}{}^m - \epsilon^{\bba\bbb\bbc\bbd\bee} \tau_{\bbb\bbc} K_{\bbd\bee}{}^m= 0 \,, \qquad 3 Z^{\bbb\bbc,\bba} R_{\bbb\bbc} - \epsilon^{\bba\bbb\bbc\bbd\bee} \tau_{\bbb\bbc} R_{\bbd\bee} = 0 \,. \label{eq:IIAReq}
\end{equation}

For completeness let us also consider the analogous consistency equations for the IIB theory. In terms of
\begin{equation}
 E_{\bba\bbb}{}^{mn} = \epsilon^{mnp} K_{p,\bba\bbb} \,, \qquad E_{\bba\bbb}{}^{45} = \frac{1}{3!} \epsilon_{mnp} R_{\bba\bbb}{}^{mnp} \,, \qquad E_{\bba\bbb}{}^{m\,u} = L_{\bba\bbb}{}^{m\,u} \,,
\end{equation}
where $u = 4, 5$ labels the $SL(2)$ symmetry of IIB, equation \eqref{eq:10CC} becomes
\begin{equation}
 \begin{split}
  - \tau_{\bba\bbb,\bbc\bbd}{}^{\bee\bbf} K_{m,\bee\bbf} &= {\cal L}_{\bba\bbb} K_{m,\bbc\bbd} = K_{n,\bba\bbb} \partial^n K_{m,\bbc\bbd} - K_{n,\bbc\bbd} \partial^n K_{m,\bba\bbb} \,, \\
  - \tau_{\bba\bbb,\bbc\bbd}{}^{\bee\bbf} R_{\bee\bbf}{}^{mnp} &= {\cal L}_{\bba\bbb} R_{\bbc\bbd}{}^{mnp} + 6 \epsilon_{uv} L_{\bbc\bbd}{}^{[m|u|} \partial^{n} L_{\bba\bbb}{}^{p]v} \,, \\
  - \tau_{\bba\bbb,\bbc\bbd}{}^{\bee\bbf} L_{\bee\bbf}{}^{m\,u} &= {\cal L}_{\bba\bbb} L_{\bbc\bbd}{}^{m\,u} + 2 K_{n,\bbc\bbd} \partial^{[m} L_{\bba\bbb}{}^{n]u} \,. \label{eq:10CCIIB}
 \end{split}
\end{equation}
Here ${\cal L}_{\bba\bbb}$ denotes the standard IIB Lie derivative, i.e. with upside-down indices (see for example \cite{Blair:2013gqa}), with the diffeomorphism parameter $K_{i,\bba\bbb}$. We see that the right-hand side of the first equation is antisymmetric under the exchange of the pair of indices $\left[\bba\bbb\right] \leftrightarrow \left[\bbc\bbd\right]$. Thus, we find that for a gauging to be of IIB origin, we must have
\begin{equation}
  3 Z^{\bbb\bbc,\bba} K_{m, \bbb\bbc} - \epsilon^{\bba\bbb\bbc\bbd\bee} \tau_{\bbb\bbc} K_{m,\bbd\bee} = 0 \,. \label{eq:IIBReq}
\end{equation}

Let us now return to the question of whether the ${\mathbf{10}'} \subset {\mathbf{40'}}$ can come from IIA. To differentiate between the IIA and IIB theories we require dependence on all three internal coordinates and so we consider the case where the gaugings of the ${\mathbf{10}'}$ are not degenerate. Using \eqref{eq:SymTorsion} it is easy to show that when $\TMP^{\balpha\bbeta} = \eta^{\balpha\bbeta}$ is not degenerate, \eqref{eq:IIAReq} can only be satisfied by a vanishing twist matrix. Thus these gaugings cannot be obtained from a IIA truncation. In particular, this applies to the $SO(4)$ theory.
By the duality established above, in turn a non-degenerate $\TM_{\balpha\bbeta} = \eta_{\balpha\bbeta}$ cannot be obtained from a IIB truncation.

This is interesting in the light of the half-maximal theory, where there is a family of $SO(4)$ gaugings involving non-degenerate gaugings in both $M_{\balpha\bbeta}$ and $\tilde{M}^{\balpha\bbeta}$, i.e. in the $\mathbf{10}$ and ${\mathbf{10}'}$ \cite{Dibitetto:2012rk}. The result here suggests that such gaugings can only be obtained by violating the section condition, as the corresponding twist matrix would be required to depend both on IIA coordinates and their dual IIB ones. Indeed, this has been shown for the half-maximal theory in \cite{Dibitetto:2012rk,Lee:2015xga}.

\section{Dualising the $\mathbf{4}$'s} \label{s:4s}
Recall from \eqref{split3} that the embedding tensor also contains two $\mathbf{4}$'s and one $\mathbf{4'}$. Can the duality discussed above be extended to these gaugings? Let us begin by relaxing the Ansatz \eqref{eq:DiagonalTwist} in order to have non-zero $\mathbf{4}$'s. Consider first
\begin{equation}
 U_a{}^{\bba} = \begin{pmatrix}
  \omega^{-1/2} V_\alpha{}^{\balpha} & \omega^{-1/2} A_\alpha \\ 0 & \omega^2
  \end{pmatrix} \,. \label{eq:UpperTriTwist}
\end{equation}
The consistency equations are then
\begin{equation}
 \begin{split}
  \TM_{\balpha\bbeta} &= \rho^{-1} \omega V_{(\balpha}{}^\alpha \partial_{|\alpha\beta|} V_{\bbeta)}{}^\beta \,, \qquad \TMP^{\balpha\bbeta} = \rho^{-1} \omega V_\alpha{}^{(\balpha} \partial^{|\alpha\beta|} V_\beta{}^{\bbeta)} \,, \\
  2 \tau_{\balpha\bbeta} &= - \rho^{-1} \omega \left( \partial_{\alpha\beta} V_{\balpha\bbeta}{}^{\alpha\beta} -5 V_{\balpha\bbeta}{}^{\alpha\beta} \partial_{\alpha\beta} \ln \omega + 6 V_{\balpha\bbeta}{}^{\alpha\beta} \partial_{\alpha\beta} \ln \left(\rho^{-1}\omega\right) \right)\,, \\
  2\Six^{\balpha\bbeta} &= \rho^{-1} \omega \left( \partial^{\alpha\beta} V_{\alpha\beta}{}^{\balpha\bbeta} - 5 V_{\alpha\beta}{}^{\balpha\bbeta} \partial^{\alpha\beta} \ln \omega \right) \,, \\
  Z^{\balpha\bfi,\bfi} &= \rho^{-1} \omega V_{\alpha\beta}{}^{\balpha\bbeta} \partial^{\alpha\beta} A_{\bbeta} + \left( \TMP^{\balpha\bbeta} + \Six^{\balpha\bbeta} \right) A_{\bbeta} \,, \label{eq:CC10+6+4}
 \end{split}
\end{equation}
where $A_{\balpha} = V_{\balpha}{}^\alpha A_\alpha$. We see that the equations for the $\mathbf{10}$'s and $\mathbf{6}$'s are unchanged but additionally the $\mathbf{4'} \subset \mathbf{40'}$ can be gauged. If we instead take the Ansatz
\begin{equation}
 U_a{}^{\bba} = \begin{pmatrix}
   \omega^{-1/2} V_\alpha{}^{\balpha} & 0 \\
   \omega^2 B^{\balpha} & \omega^2
  \end{pmatrix}\,, \label{eq:LowerTriTwist}
\end{equation}
we again find the same $\mathbf{10}$'s and $\mathbf{6}$'s as in \eqref{eq:CC10+6} but additionally the following can be gauged:
\begin{equation}
 \begin{split}
  \tau_{\balpha\bfi} &= - \frac{1}{2} \rho^{-1} \omega V_{\balpha\bbeta}{}^{\alpha\beta} \partial_{\alpha\beta} B^{\bbeta} - B^{\bbeta} \tau_{\balpha\bbeta} \,, \\
  Z^{\balpha\bbeta,\bgamma} &= \frac{1}{2} \rho^{-1} \omega \epsilon^{\balpha\bbeta\bdelta\brho} V_{\bdelta\brho}{}^{\delta\rho} \partial_{\delta\rho} B^{\bgamma} - \frac{2}{3} \epsilon^{\balpha\bbeta\bgamma\brho} \left( S_{\brho\bfi} + B^{\bdelta} \TM_{\brho\bdelta} \right) + 2 B^{[\balpha} \left( \TMP^{\bbeta]\bgamma} + \Six^{\bbeta]\bgamma} \right) - 2 B ^{[\balpha} \Six^{\bbeta\bgamma]} \,, \\
  S_{\balpha \bfi} &= - \frac{1}{2} \rho^{-1} \omega V_{\balpha\bbeta}{}^{\alpha\beta} \partial_{\alpha\beta} B^{\bbeta} - B^{\bbeta} \TM_{\balpha\bbeta} \,, \\
  S_{\bfi\bfi} &= - B^{\balpha} \left( 2 S_{\balpha\bfi} + B^{\bbeta} \TM_{\balpha\bbeta} \right) \,. \label{eq:CCb4}
 \end{split}
\end{equation}
The $SL(4)$ (co-)vectors $A_\alpha$ and $B^\alpha$ should be exchanged by the outer automorphism of $SL(4)$ so that
\begin{equation}
 V_\alpha{}^{\balpha} \longleftrightarrow \left(V_{\balpha}{}^{\alpha}\right)^{-T} \,, \qquad \partial_{\alpha\beta} \longleftrightarrow \partial^{\alpha\beta} \,, \qquad A_\alpha \longleftrightarrow B^\alpha \,. \label{eq:Aut4s}
\end{equation}
This maps a solution of the equations \eqref{eq:CCb4} to a solution of \eqref{eq:CC10+6+4} but not vice versa. Thus, it is not in general possible to map a twist that gauges the $\mathbf{4'} \subset \mathbf{40'}$ into a twist gauging the $\mathbf{4} \subset \mathbf{15}$, $\mathbf{4} \subset \mathbf{10}$ and $\mathbf{20'} \subset \mathbf{40'}$. Furthermore, if we start with a gauging of the $\mathbf{4'} \subset \mathbf{40'}$ that satisfies the quadratic constraints \eqref{eq:QC} and perform the duality to obtain a gauging in the $\mathbf{4} \subset \mathbf{15}$, $\mathbf{4} \subset \mathbf{10}$ and $\mathbf{20'} \subset \mathbf{40'}$, then this dual gauging does not in general satisfy the quadratic constraint. Then the dual gaugings do not define a consistent gauged SUGRA. We will see an example of this in section \ref{s:4sEx}.

\section{Further examples} \label{s:OtherEx}

We will now use our twist Ans\"atze \eqref{eq:DiagonalTwist}, \eqref{eq:UpperTriTwist} and \eqref{eq:LowerTriTwist} and the duality discussed above to obtain new uplifts of various maximal gauged SUGRAs. This is not an exhaustive list of solutions to the quadratic constraints, but rather a selection of examples for which uplifts to type II SUGRA can be constructed nicely with the twist Ans\"atze we have considered so far. The gaugings we consider are summarised in table \ref{t:1}. Each value of $\alpha$ in the range $-\pi/2 \leq \alpha \leq \pi$, as well as each $\lambda$ taking the values $\lambda = 1,\, \frac12,\, 0$, each $\eta,\, \eta' = \pm 1$ and each $a \in \mathbb{R}$ labels different inequivalent orbits.
Note that for orbits 1 and 7 -- 9 we have indicated that the gaugings in the $\mathbf{4}$ vanish. This is because any non-zero gaugings in the $\mathbf{4}$ allowed by the quadratic constraint \eqref{eq:QC} can be removed by an $SL(5)$ transformation and thus lead to equivalent 7-dimensional theories. Orbits 6 -- 9 involve the trombone gauging (when $\lambda \neq 0$) and thus the 7-dimensional theories they represent do not admit an action principle. We will see in section~\ref{s:10+6Ex} that in some cases their uplifts are non-geometric, where the trombone scaling symmetry is used to patch together the solution.

\bigskip
\noindent\makebox[\textwidth]{
 \begin{minipage}{\textwidth}
  \begin{center}
  \begin{tabular}{|c|c|c|c|c|c|}
   \hline
   Orbit & $\TM_{\balpha\bbeta}$ & $\TMP^{\balpha\bbeta}$ & $Z^{\balpha\bfi,\bfi}$ & $\Six^{\bar{2}\bar{3}}$ & $\tau_{\bar{1}\bar{4}}$ \Tstrut\Bstrut \\ \hline
   1 & $\diag \left(\eta, 1, 0, 0\right)$ & $ \diag \left(0, 0, \eta, 1\right)$ & 0 & 0 & 0 \\
   2 & $\diag \left(1, 0, 0, 0\right)\cos\alpha$ & $\diag \left(0, 0, 0, 1\right) \sin\alpha$ & 0 & 0 & 0 \\
   3 & $\diag \left(\eta, \eta', 1, 0\right)$ & 0 & $\left(0, 0, 0, 1\right)$ & 0 &  \\
   4 & $\diag \left(\eta, 1, 0, 0\right)$ & 0 & $\left(0, 0, 0, 1\right)$ & 0 & 0 \\
   5 & $\diag \left(1, 0, 0, 0\right)\cos\alpha$ & $\diag \left(0, 0, 0, 1\right) \sin\alpha$ & $\left(0,0,1,0\right)$ & 0 & 0 \\
   6 & $\diag\left(0, 0, 0, 0\right)$ & $\diag\left(0, 0, 0, 0\right)$ & 0 & $\lambda - 1$ & $\lambda$ \\
   7 & $\diag \left(\eta, 1, 0, 0\right)$ & 0 & 0 & $-a$ & $a$ \\
   8 & $\diag \left(\eta, 1, 0, 0\right)$ & $ \diag \left(0, 0, \eta, 1\right)$ & 0 & $-a$ & $a$ \\
   9 & $\diag \left(1, 0, 0, 0\right)\cos\alpha$ & $\diag \left(0, 0, 0, 1\right) \sin\alpha$ & 0 & $-a$ & $a$ \\
   \hline
  \end{tabular}
 \captionof{table}{\small{Orbits of gaugings for which we will construct uplifts. Each $\alpha$ in the range $-\pi/2 \leq \alpha \leq \pi$, each $\lambda = 1,\, \frac12,\, 0$, each $\eta,\,\eta' = \pm 1$ and each $a \in \mathbb{R}$ labels different inequivalent orbits.}  } \label{t:1}
  \end{center}
 \end{minipage}
}

\subsection{Orbits 1 and 2} \label{s:10+10Ex}
In section \ref{s:NoGo10s} we showed that non-degenerate gaugings in the $\mathbf{10}$ descend from IIA and those in the $\mathbf{10'}$ descend from IIB. Let us now uplift gaugings which mix the $\mathbf{10}$ and $\mathbf{10'}$. The quadratic constraint is now
\begin{equation}
 \TMP^{\balpha\bbeta} \TM_{\balpha\bgamma} = 0 \,.
\end{equation}
The solutions are given by orbits 4 -- 11 of \cite{Dibitetto:2012rk}.

\paragraph{Orbit 1}
This orbit can be represented by the gaugings
\begin{equation}
 \TM_{\balpha\bbeta} = \diag \left(\eta, 1, 0, 0 \right) \,, \qquad \TMP^{\balpha\bbeta} = \diag\left(0, 0, \eta, 1 \right) \,.
\end{equation}
These correspond to an embedding of orbits 6 and 9 (with $\alpha = \pi/4$) of \cite{Dibitetto:2012rk} into the maximal theory.

The twist matrices are given by
\begin{equation}
 \begin{split}
  V_\alpha{}^{\balpha} &= \begin{pmatrix}
   \left(1-v\right)^{1/4} & - \eta y_1 \left(1-v\right)^{-1/4} & 0 & 0 \\
   y_1 \left(1-v\right)^{-1/4} & \left(1-v\right)^{1/4} & 0 & 0 \\
   0 & 0 & \left(1-v\right)^{3/4} & - y_1 \left(1-v\right)^{1/4} \\
   0 & 0 & \left(1-v\right)^{1/4} \eta y_1 & \left(1-v\right)^{3/4}
  \end{pmatrix} \,, \\
  \omega &= \left(1-v\right)^{1/10} \,, \label{eq:10+10'1}
 \end{split}
\end{equation}
with $\rho = \omega$ and where $v = \eta y_1^2$ and $u = y_1^2$. From \eqref{eq:MParam} we find the internal space in string frame to be
\begin{equation}
 \begin{split}
  \mathring{ds}^2 &= \left(1-v\right)^{-1} dy_1^2 + dy_2^2 + dy_3^2 - 2 y_1 \left(\eta-1\right) \left(1-v\right)^{1/2} \left(1+u-v\right)^{-1} dy_2 dy_3 + ds_7^2 \,, \\
  \mathring{B}_{23} &= y_1 \left(\eta-1\right) \left(1-v\right)^{1/2} \left(1+u-v\right)^{-1} \,, \qquad e^{\mathring{\varphi}} = \left(1+u-v\right)^{-1/2} \,.
 \end{split}
\end{equation}
Note that when $\eta = 1$ the background is the Kaluza-Klein circle encountered in \eqref{eq:IIBSphereInt}. However, the internal space will be different at other points in the scalar moduli space. It is of course also possible to generate the gaugings
\begin{equation}
 \TM_{\balpha\bbeta} = \diag \left( 0, 0, \eta, 1 \right) \,, \qquad \TMP^{\balpha\bbeta} = \diag \left( \eta, 1, 0, 0 \right) \,,
\end{equation}
by applying the duality discussed in section \ref{s:Aut}. As before, the internal space remains the same under the duality.

\paragraph{Orbit 2} These orbits describe an embedding of orbits 11 of \cite{Dibitetto:2012rk} into the maximal theory. The gaugings are
\begin{equation}
 M_{\balpha\bbeta} = \diag \left(0, 1, 0, 0\right) \cos\alpha \,, \qquad \tilde{M}^{\balpha\bbeta} = \diag \left(0, 0, 0, 1 \right) \sin\alpha \,,
\end{equation}
where $ - \pi / 2 \leq \alpha \leq \pi$ gives the range of inequivalent orbits. The twist matrices are given by
\begin{equation}
 \begin{split}
  V_\alpha{}^{\balpha} &= \begin{pmatrix}
   1 & - y_1 \cos\alpha & 0 & 0 \\
   0 & 1 & 0 & 0 \\
   0 & 0 & 1 & - y_1 \sin\alpha \\
   0 & 0 & 0 & 1
  \end{pmatrix} \,, \qquad \omega = 1 \,, \label{eq:10+10'2}
 \end{split}
\end{equation}
where $y_1 = Y^{14}$ and the internal space is given by
\begin{equation}
 \begin{split}
  ds^2 &= dy_1^2 + dy_2^2 + \left(dy_3 - y_1 \sin\alpha ~ dy_2 \right)^2 + ds_7^2 \,, \\
  B_{23} &= y_1 \cos\alpha \,, \\
 \end{split}
\end{equation}
with all other fields vanishing. The dual gaugings $M_{\balpha\bbeta} \longleftrightarrow \tilde{M}^{\balpha\bbeta}$ are in this case equivalent to the gaugings discussed.

\subsection{Orbits 3 and 4} \label{s:4sEx}

When $M_{\balpha\bbeta}$ and $Z^{\balpha\bfi,\bfi}$ are the only non-zero gaugings, the quadratic constraint is
\begin{equation}
 \TM_{\balpha\bbeta} Z^{\bbeta\bfi,\bfi} = 0 \,. \label{eq:4QC}
\end{equation}
Thus, $Z^{\balpha\bfi,\bfi} \neq 0$ only when $M_{\balpha\bbeta}$ is degenerate. Let us consider separately the cases where $\TM_{\balpha\bbeta}$ has rank $3$ and rank $2$, corresponding to orbits 3 and 4 in table \ref{t:1}, respectively.

\paragraph{Orbit 3}
Take $\TM_{\bar{\imath}\bar{\jmath}} = \eta_{\bar{\imath}\bar{\jmath}}$, $M_{\bfo\bfo} = 1$ and $\TM_{\bar{3}\bar{3}} = 0$, with $\bar{\imath}, \bar{\jmath} = 1, 2$ and all other elements vanishing. Then by \eqref{eq:4QC} we can only have
\begin{equation}
 Z^{\bar{3}\bfi,\bfi} = c \,.
\end{equation}
We could use an $SL(5)$ transformation to set $c = 1$ but we will not do so here to keep track of $c$ in the internal space. However, the reader should keep in mind that all values of $c \neq 0$ correspond to equivalent 7-dimensional theories.

From the no-go theorem \eqref{eq:IIAReq} one finds that this gauging cannot be obtained by a IIA truncation. It can, however, be lifted to 10-dimensional IIB SUGRA using the Ansatz \eqref{eq:UpperTriTwist} with the same $V_{\alpha}{}^{\balpha}$ as in \eqref{eq:IIBSphere} with $r = 1$ and with
\begin{equation}
 A_4 = - c y_3 \left(1-v\right)^{-1/4} \,, \qquad A_i = A_3 = 0 \,,
\end{equation}
where $y_3 = Y^{12}$ is the third IIB coordinate. Recall that the other two coordinate are given by $y_1 = Y^{14}$, $y_2 = Y^{24}$. The background for this truncation is given by
\begin{equation}
 \begin{split}
  \mathring{ds}^2 &= ds_7^2 + \left(1-v\right)^{-1} \left[\delta^{ij} dy_i dy_j - \frac{\left(\eta^{ij} y_i dy_j\right)^2}{1+u-v} \right] \\
  & \quad + \left(1+u-v\right) \left[ dy_3 + \left(1-v\right)^{-1/2} \frac{ 1 + K \left(1+u-v\right) }{1+u-v} \eta^{ij} y_i dy_j \right]^2 \,, \\
  \mathring{C}^{ij} &= - c y_3 \left(1-v\right)^{-1/2} \epsilon^{ij} \,. \label{eq:IIB4}
 \end{split}
\end{equation}
As before, we use the convention of \cite{Blair:2013gqa} where IIB indices are placed ``upside-down'' and $\mathring{C}^{ij}$ labels the Ramond-Ramond two-form. The metric here is the T-dual of the $H^{p,q}$ solutions in \eqref{eq:IIBSphereInt}. Furthermore, only the two-form depends on $c$ and the NS-NS sector remains invariant as $c$ is turned on.

\paragraph{Orbit 4} Take $\TM_{\bar{1}\bar{1}} = \eta$, $\TM_{\bar{4}\bar{4}} = 1$ and all other components vanishing. Then by \eqref{eq:4QC} we can have the gaugings
\begin{equation}
 Z^{\bar{2}\bfi,\bfi} = c \,, \qquad Z^{\bar{3}\bfi,\bfi} = d \,.
\end{equation}
One can use an $SL(5)$ transformation to set $c = 0$ and $d = 1$ but we will not do so here to keep track of where the gaugings appear in the internal space. Once again, however, the reader should remember that different values of $c$ and $d$ (with at least one non-vanishing) correspond to the same 7-dimensional theory.

We again use the Ansatz \eqref{eq:UpperTriTwist} with $V_\alpha{}^{\balpha}$ as in \eqref{eq:IIBSphere} with $r = 2$ and solve the gauging of the $Z^{\balpha\bfi,\bfi}$ by
\begin{equation}
 A_1 = \left(1-v\right)^{-1/4} \left(cy_3 - dy_2\right) \,,
\end{equation}
with all other $A_\alpha = 0$, $\alpha \neq 1$. The twist now only depends on $y_1 = Y^{14}$, $y_2 = Y^{24}$ and $y_3 = Y^{34}$ and so gives an uplift to IIA supergravity. From \eqref{eq:MParam} the internal space is found to be
\begin{equation}
 \begin{split}
  \mathring{ds}_{11}^2 &= \left(1+u-v\right)^{-2/3} \left[ dy_2^2 + dy_3^2 + \left( dz + C_1 dy_1 \right)^2 \right] \\
  & \quad + \left(1+u-v\right)^{1/3} \frac{dy_1^2}{1-v} + \left(1+u-v\right)^{-2/3} ds_7^2 \,, \\
   \mathring{C}_{23z} &= \left(1+u-v\right)^{-1} \left(1-v\right)^{1/2} \left(1-\eta\right) y_1 \,, \\
   \mathring{C}_1 &= \left(1-v\right)^{-1/2} \left(cy_3 - dy_2 \right) \,. \label{eq:IIA4}
 \end{split}
\end{equation}
This is the same circle / hyperbola reduction as in \eqref{eq:IIBSphereInt} but with an additional Ramond-Ramond one-form $\mathring{C}_1$ turned on. Similar to orbit 3, only the Ramond-Ramond one-form depends on $c$ and $d$.

To conclude the discussion of these orbits, let us consider the dual gaugings. The duality would give gaugings of the $\mathbf{4} \subset \mathbf{15}$, $\mathbf{4} \subset \mathbf{10}$ with
\begin{equation}
 S_{\balpha 5} = \tau_{\balpha 5} \,,
\end{equation}
as well as possibly the $\mathbf{20}'$. However, these gaugings violate the quadratic constraint \eqref{eq:QC} and hence they do not define a consistent gauged SUGRA.

\subsection{Orbit 5} \label{s:10+4'Ex}
For the gaugings $M_{\balpha\bbeta} = \diag\left(0, 1, 0, 0\right) \cos\alpha$ and $\tilde{M}^{\balpha\bbeta} = \diag\left(0,0,0,1\right) \sin\alpha$ the quadratic constraint allows the $\mathbf{4}$'s
\begin{equation}
 Z^{\balpha\bfi,\bfi} = \left(d, 0, c, e\right) \,.
\end{equation}
We can use an $SL(5)$ transformation to make two of these vanish and scale the third. Let us thus take $d = e = 0$ but keep $c \neq 1$ in general so that we can see where it ends up in the internal space. The twist matrix is then given by Ansatz \eqref{eq:UpperTriTwist} with $V_\alpha{}^{\balpha}$ as in \eqref{eq:10+10'2} and
\begin{equation}
 A_{\bar{3}} = c y_1 \,.
\end{equation}
The internal space is then given by
\begin{equation}
 \begin{split}
  \mathring{ds}^2 &= dy_1^2 + dy_2^2 + \left(dy_3 - y_1 \sin\alpha~ dy_2 \right)^2 + ds_7^2 \,, \\
  \mathring{B}_{23} &= y_1 \cos\alpha \,, \\
  \mathring{C}_2 &= - c y_1^2 \sin\alpha \,, \\
  \mathring{C}_3 &= c y_1 \,.  
 \end{split}
\end{equation}
As for orbits 3 and 4 we find that the parameter $c$ only appears in the Ramond-Ramond 1-form. The dual gaugings would again not satisfy the quadratic constraint \eqref{eq:QC}.

\subsection{Orbit 6} \label{s:6sEx}
To keep our formulae simple we will actually uplift the gaugings
\begin{equation}
 \tau_{\bar{1}\bar{4}} = 3 \left(\lambda - 1\right) \,, \qquad \Six^{\bar{2}\bar{3}} = 3 \lambda \,,
\end{equation}
with inequivalent gaugings for $\lambda = 1, \frac12, 0$. We can obtain these gaugings easily using the block-diagonal Ansatz for the twist matrix \eqref{eq:DiagonalTwist} and by choosing the scalars $\rho$ and $\omega$ appropriately.

The twist matrix is given by $V_\alpha{}^{\balpha} = \delta_\alpha{}^{\balpha}$ with scalars $\omega = \left(1 - y_1 \right)^{6\lambda/5}$ and $\rho = \left(1-a \cdot y\right)^{6\lambda/5 - 1}$. The internal space in string frame is
\begin{equation}
 \mathring{ds}^2 = dy_m dy^m + \left(1-a\cdot y\right)^2 ds_7^2 \,, \qquad e^{\mathring{\varphi}} = \left(1-a \cdot y\right)^{3\lambda} \,.
\end{equation}
We can see that the string-frame metric is independent of $\lambda$ and the dilaton tunes between the different gaugings. In particular, when $\lambda = 1$ we have a standard 7-dimensional gauged SUGRA, whereas for the cases $\lambda = 0$ and $\lambda = 1/2$ the 7-dimensional theory does not have an action principle, even though it can still be uplifted to 10-dimensional SUGRA. For each $\lambda$ the outer automorphism discussed in section \ref{s:Aut} relates equivalent gaugings.

\subsection{Orbits 7 -- 9} \label{s:10+6Ex}

The gaugings we consider here involve some of the gaugings encountered previously in this paper together with both $\mathbf{6}$'s. These can be uplifted by using almost the same twist matrices as without the $\mathbf{6}$'s. In particular we will keep $V_\alpha{}^{\balpha}$ unchanged but change $\rho = \omega$. Let us write $\rho = \omega = \omega_0 h$, where $\omega_0$ is the value of $\omega$ where the $\mathbf{6}$'s vanish. The function $h$ then has to satisfy
\begin{equation}
 2 \tau_{\balpha\bbeta} = - 2 \Six_{\balpha\bbeta} = 5 V_{\balpha\bbeta}{}^{\alpha\beta} \partial_{\alpha\beta} \ln h \,. \label{eq:hConsistency}
\end{equation}

\paragraph{Orbit 7}
Let us start with the IIA sphere / hyperboloid case \eqref{eq:IIASphere} where $\omega_0 = \left(1-v\right)^{1/10}$. The quadratic constraint \eqref{eq:QC} implies the only gaugings with non-zero $\Six^{\balpha\bbeta} = - \tau^{\balpha\bbeta}$ are given by
\begin{equation}
 M_{\balpha\bbeta} = \diag\left(\eta, 1, 0, 0\right) \,, \qquad \tau_{\bar{1}\bar{4}} = - \Six^{\bar{2}\bar{3}} = a \,.
\end{equation}
For $a = 0$ these are $S^1$ and $H^1$ reductions. Now, we find
\begin{equation}
 h = \exp\left( \frac{2a \arcsin\left(\sqrt{\eta}y_1\right)}{5\sqrt{\eta}} \right) \,. \label{eq:hCircle}
\end{equation}
The internal space in string-frame is given by
\begin{equation}
 \begin{split}
  \mathring{ds}^2 &= \frac1{1-v} dy_1^2 + \frac1{1+u-v}\left(dy_2^2 + dy_3^2\right) + ds_7^2 \,, \\
  \mathring{B}_{23} &= \left(1-v\right)^{1/2} \left(1+u-v\right) \left(\eta - 1 \right) y_1 \,, \\
  e^{\mathring{\varphi}} &= \left(1+u-v\right)^{-1/2} \exp\left(\frac{a\arcsin\left(\sqrt{\eta}y_1\right)}{\sqrt{\eta}}\right) \,.
 \end{split}
\end{equation}
We see that when $\eta = 1$, the internal space is non-geometric because the dilaton is not globally well-defined. Instead, it is patched by the trombone scaling symmetry of the equations of motion. This is a reminiscent of the non-geometric construction in \cite{Shahbazi:2015sba} albeit in seven dimensions.

\paragraph{Orbit 8} For the gaugings
\begin{equation}
 M_{\balpha\bbeta} = \diag\left(\eta,1,0,0\right) \,, \qquad \tilde{M}^{\balpha\bbeta} = \diag\left(0,0,\eta,1\right) \,, \qquad \tau_{\bar{1}\bar{4}} = - \Six^{\bar{2}\bar{3}} = a \,,
\end{equation}
with $V_\alpha{}^{\balpha}$ and $\omega$ as in \eqref{eq:10+10'1}, the consistency condition on $h$, \eqref{eq:hConsistency}, has the same solution $h$ as in \eqref{eq:hCircle}. We find the internal space in string-frame
\begin{equation}
 \begin{split}
  \mathring{ds}^2 &= \left(1-v\right)^{-1} dy_1^2 + dy_2^2 + dy_3^2 + 2 y \left(\eta-1\right) \left(1-v\right)^{1/2} \left(1+u-v\right)^{-1} dy_2 dy_3 + ds_7^2 \,, \\
  \mathring{B}_{23} &= \left(1-v\right)^{1/2} \left(1+u-v\right)^{-1} y \left(\eta-1\right) \,, \\
  e^{\mathring{\varphi}} &=  \left(1+u-v\right)^{-1/2} \exp\left(\frac{a\arcsin\left(\sqrt{\eta}y_1\right)}{\sqrt{\eta}} \right) \,.
 \end{split}
\end{equation}

\paragraph{Orbit 9} For the gaugings
\begin{equation}
 M_{\balpha\bbeta} = \diag \left( 0, 1, 0, 0 \right) \cos\alpha \,, \qquad \tilde{M}^{\balpha\bbeta} = \diag \left( 0, 0, 0, 1 \right) \sin\alpha \,, \qquad \tau_{\bar{1}\bar{4}} = - \Six^{\bar{2}\bar{3}} = a \,,
\end{equation}
with $V_\alpha{}^{\balpha}$ and $\omega_0$ as in \eqref{eq:10+10'2} we find
\begin{equation}
 h = \exp \left( \frac{2a}{5} y_1 \right) \,.
\end{equation}
The internal space in string-frame is
\begin{equation}
 \begin{split}
  \mathring{ds}^2 &= dy_1^2 + dy_2^2 + \left(dy_3 - y_1 \sin\alpha~ dy_2 \right)^2 + ds_7^2 \,, \\
  \mathring{B}_{23} &= y_1 \cos\alpha \,, \\
  e^{\mathring{\varphi}} &= \exp\left(a y_1 \right) \,.
 \end{split}
\end{equation}

\section{Conclusions} \label{s:Conclusion}
In this paper we studied consistent truncations of type IIA and IIB SUGRA to 7-dimensional maximal gauged SUGRA using exceptional field theory. By using a $GL(4)$ Ansatz for the twist matrices, we showed that IIA / IIB consistent truncations are related by the outer automorphism of $SL(4)$ which acts on the irreducible representations of the embedding tensor as
\begin{equation}
 \mathbf{10} \longleftrightarrow \mathbf{10}' \,, \qquad \mathbf{6}_{40} \longleftrightarrow \mathbf{6}_{40} \,, \qquad \mathbf{6}_{10} \longleftrightarrow \mathbf{6}_{10} \,.
\end{equation}
Here $\mathbf{6}_{40}$ and $\mathbf{6}_{10}$ denote the $\mathbf{6}$'s coming from the $\mathbf{40}'$ of $SL(5)$ and from the trombone gauging, respectively. We also showed that this duality between IIA and IIB consistent truncations always exists when the embedding tensor has vanishing components in the $\mathbf{4}'$ of $SL(4)$. Otherwise, the dual gaugings will in general not satisfy the quadratic constraints \eqref{eq:QC}.

We used this duality to prove the consistent truncation of IIB on $S^3$ and $H^{p,q}$ by constructing twist matrices that give rise to the relevant $CSO(p,q,r)$ gaugings with embedding tensor in the $\mathbf{40}'$. The twist matrices are dual to those describing the IIA uplift of gaugings in the ${\bf 15}$±\cite{Hohm:2014qga}. Using the dictionary between EFT and IIA / IIB fields, we used the twist matrices to derive the full truncation Ans\"atze for the internal sectors of the IIA and IIB reductions. They were shown to coincide in the NS-NS sector. This is a general feature of the duality: it relates truncations with the same NS-NS sector. Finally, from the form of the consistency equations we derived some no-go theorems showing that non-degenerate gaugings with IIA origin cannot also be uplifted to IIB and vice versa.

In the second part of this paper we further generalised the twist matrices of \cite{Hohm:2014qga} to uplift other gaugings of 7-dimensional maximal gauged SUGRA to type II SUGRA. These examples include gaugings of the $\mathbf{15}$ and $\mathbf{40}'$ simultaneously, and of the trombone, where the gauged SUGRA does not admit a Lagrangian. In the latter case, the internal space of the truncation is only well-defined up to the $\mathbb{R}^+$ scaling symmetry of the equations of motion. Among the direct applications of these uplift formulas is the higher-dimensional embedding of the vacua found in the lower-dimensional theories, such as \cite{Dibitetto:2015bia}.

The twist matrices used throughout this paper are defined in local patches. For the truncation to be consistent, these twist matrices must yield a generalised parallelisation \cite{Lee:2014mla}. To show this we would have to patch our twist matrices to obtain globally well-defined vector fields. A patching prescription for exceptional field theory is still lacking, although it is known for double field theory \cite{Hohm:2012gk,Park:2013mpa,Berman:2014jba,Naseer:2015tia}. Whatever this covariant patching prescription will be, it should consist of the global $SL(5) \times \mathbb{R}^+$ symmetries of the 7-dimensional SUGRA. We can thus argue that our twist matrices are well-defined by checking that the internal space they define is well-defined up to $SL(5) \times \mathbb{R}^+$ dualities. This is indeed the case for all the examples given here.

The duality established in this paper exchanges IIA and IIB consistent truncations, by relating
different irreducible representations of the embedding tensor of 7-dimensional gauged supergravity according to the embedding (\ref{split3}). Similar dualities are expected to arise in all dimensions.
In contrast to the 7-dimensional case, for all other dimensions the embedding tensor $X_{ABC}$ of the half-maximal
theory sits in an irreducible representation of $SO(d,d)$, thus in a single irreducible representation of the 
$E_{d+1(d+1)}$ duality group of the maximal theory. It is thus less clear if the resulting gaugings sit in different orbits 
of the duality group according to their IIA / IIB origin, i.e.\ if IIA and IIB reductions give rise to inequivalent lower-dimensional theories.
A natural starting point for further investigation are 3-dimensional maximal gauged SUGRAs. These are known to have two inequivalent $SO(8)$ gaugings, expected to arise from $S^7$ reductions of IIA / IIB \cite{Fischbacher:2003yw}. Indeed, the full EFT has been constructed for this case \cite{Hohm:2014fxa} so that the full reduction Ans\"atze of the $S^7$ truncations could then also be derived.
It would also be interesting to cast into this framework consistent truncations of the massive IIA theory such as \cite{Guarino:2015jca} which would require a (modest) dependence of the twist matrices on one of the non-physical coordinates, c.f.~\cite{Hohm:2011cp}.

Finally, it would be interesting to try and find a systematic procedure for the construction of the twist matrices for all possible allowed gaugings of the quadratic constraint \eqref{eq:QC}. An interesting proposal for the case of half-maximal gauged SUGRA appeared in \cite{Bosque:2015jda}. However, the resulting twist matrices are not $O(d,d)$-valued so that it is not immediately clear how to find the associated reduction Ans\"atze.

\section{Acknowledgements}
We would like to thank Michael Abbott, Charles Strickland-Constable, Olaf Hohm, Diego Marqu\'{e}s, Jose Fern\'{a}ndez-Melgarejo and Alejandro Rosabal for helpful discussions. EM would like to thank the organisers of the Simons Summer Workshop 2015 and the CERN-CKC TH Institute on Duality Symmetries in String and M-Theories for hospitality, while part of this work was being completed. EM is supported by the ERC Advanced Grant ``Strings and Gravity" (Grant No. 32004) and was also funded by the National Research Foundation (NRF) of South Africa under grant CSUR13091742207 during the initial stages of this project.


\providecommand{\href}[2]{#2}\begingroup\raggedright\endgroup

\end{document}